# A first-order magnetic phase transition near 15 K with novel magnetic-field-induced effects in $Er_5Si_3$

Niharika Mohapatra[1, 2, *], K Mukherjee[2], Kartik K Iyer[2] and E. V. Sampathkumaran[2]

[1]School of Basics Sciences, Indian Institute of Technology-Bhubaneswar, Bhubaneswar - 751013, India
[2]Tata Institute of Fundamental Research, Homi Bhabha Road, Mumbai – 400005, India

**Abstract**

We present magnetic characterization of a binary rare-earth intermetallic compound $Er_5Si_3$, crystallizing in $Mn_5Si_3$–type hexagonal structure, through magnetization, heat capacity, electrical resistivity and magnetoresistance measurements. Our investigations confirm that the compound exhibits two magnetic transitions with decreasing temperature, first one at 35 K and the second one at 15 K. The present results reveal that the second magnetic transition is a disorder-broadened first-order transition, as shown by thermal hysteresis in the measured data. Another important finding is that, below 15 K, there is a magnetic-field-induced transition with a hysteretic effect with the electrical resistance getting unusually enhanced at this transition and the magnetoresistance (MR) is found to exhibit intriguing magnetic-field dependence indicating novel magnetic phase-co-existence phenomenon. It thus appears that this compound is characterized by interesting magnetic anomalies in the temperature-magnetic field phase diagram.

### I. Introduction

Compounds with spiral spin structure, in which the angle of spins with respect to the propagation vector varies periodically; have drawn a special attention in recent years due to the strong coupling of the magnetic degrees of freedom to the lattice and electronic degrees of freedom. Among the insulating systems, such as $TbMnO_3$, $TbMn_2O_5$, $Ni_3V_2O_8$ and $CoCr_2O_4$, the effect of these coupling is manifested in terms of giant magneto-electric effect and magnetoelastic coupling [1-3]. The metallic systems with spiral magnetism also exhibit a range of unusual behavior such as quantum phase transition [4], giant magnetoresistance [5,6] and giant magnetocaloric effect [7], due to this cross-coupling. Spiral magnetic order essentially arises as a result of two or more competing magnetic exchange interactions, for example, nearest neighbor and next-nearest neighbor interactions. Therefore the intriguing physical properties of these systems are very sensitive to the external field, pressure and doping, which often disturb the subtle balance in the correlation leading to novel effects. It is therefore of interest to investigate the materials with complex magnetic structures in depth.

With this motivation, in this article, we focus on a compound $Er_5Si_3$, known to crystallize in $Mn_5Si_3$ – type hexagonal structure (space group: $P6_3/mcm$). In this structure, Er occupies two inequivalent crystallographic sites, viz., one at the tetrahedral site (4d) and the other at the octahedral site (6g), while Si is at the octahedral site only (6g). Due to the presence of two different chemical environments for the two Er atoms, a complex spiral antiferromagnetic (AFM) order sets in at low temperatures. The compound has been reported [8] to exhibit two successive (spiral) antiferromagnetic transitions with decreasing temperature, the first one at 29 K and the other at 13 K with different propagation vectors and it was claimed through neutron diffraction results that there is a narrow temperature interval around 13 K in which two magnetic phases coexist the reason for which was not obvious. Here, we report an extensive study of this compound through electrical resistivity ($\rho$), heat capacity ($C$) and dc magnetization ($M$) measurements. Though our results confirm the existence of two magnetic transitions, these occur at marginally higher temperatures, viz. 35 K and 15 K respectively. We show that the second magnetic transition near 15 K is a disorder-broadened first-order magnetic transition, as revealed by hysteresis in the measured data around this transition, thereby resolving the origin of magnetic phase-coexistence inferred in Ref.8. Another noteworthy finding we report here is that, in the magnetic-field ($H$) – temperature ($T$) plane below 15 K, there is a field induced transition, with hysteretic effects, and the shapes of magnetoresistance (MR) versus $H$ plots are rather unique indicating novel magnetic phase-co-existence phenomenon in the $H$-$T$ phase diagram as well.

### II. Experimental details

Polycrystalline sample of $Er_5Si_3$ was prepared by conventional arc-melting stoichiometric amounts of high purity Er (99.9 wt %) and Si (99.99 wt %) in argon atmosphere. For better homogeneity, the ingot was re-melted several times by flipping over each time. The weight-loss after the final melting was found to be less than 1%. The sample thus obtained was characterized by X-ray diffraction (XRD) and scanning electron microscope (SEM) and found to be single phase. The room temperature XRD lines could be indexed to $Mn_5Si_3$–type hexagonal structure with the lattice constants, $a = 8.25$ Å and $c = 6.236$ Å. These values are close to those reported earlier for $Er_5Si_3$ [9]. Dc magnetization measurements in presence of several magnetic fields (100 Oe, 5 kOe, 10 kOe, 30 kOe) as a function of temperature (1.8 – 300 K) were performed with the help of a commercial (Quantum Design) superconducting quantum interference device (SQUID) as well as vibrating sample magnetometer (VSM) (Oxford Instruments). The same magnetometers were employed to perform isothermal magnetization measurements at several temperatures. The ρ measurements (1.8–300 K) were carried out in the presence of magnetic fields by a commercial (Quantum Design) physical property measurements system (PPMS) by a conventional four-probe method. The electrical contacts of the leads to the specimen were made by a conducting silver paste. In order to identify magnetic transitions, we have also performed $C$ measurements (1.8–150 K) with the same PPMS by the relaxation method.

### III.    Results and discussion
### III A: Temperature dependence of magnetization

Figure 1 shows the temperature dependence of dc *M/H* measured in the presence of various fields. The magnetic susceptibility ($\chi$) follows Curie-Weiss behavior above 100 K as shown in the inset of Fig. 1. The linear fit of $\chi^{-1}$ yields a positive paramagnetic Curie temperature ($\theta_P = 38$ K) suggesting dominance of ferromagnetic interactions, though the net observed magnetic structure is antiferromagnetic-like. The effective moment obtained from the Curie-Weiss fit turns out to be 9.6 $\mu_B$/Er which is very close to the theoretical value for the trivalent Er. In order to understand the magnetically ordered state better, the magnetization was measured under zero-field-cooled (ZFC, from 100 K), field-cooled-cooling (FCC) and field-cooled warming (FCW) conditions at several fields. The behavior for ZFC, FCC and FCW protocols are quite different for different strengths of applied fields. In the data taken with low fields (100 Oe), there is a distinct evidence for two magnetic transitions at 35 and 15 K under all the three measurement conditions. These values of the transition temperature are marginally higher than that reported from neutron diffraction results [8]. Possibly, the transition temperatures could be sample-dependent depending sensitively on the stoichiometry and heat-treatment.  A point of emphasis is that there is in fact a thermal hysteresis in a narrow range below 15 K as shown in figure 1. This is a signature of disorder-induced first-order transition [10] and therefore the reported phase co-existence in the range 10-13 K is a result of this phenomenon.  The difference in FCC and FCW curves were observed in the *T* range 10 – 15 K for the applied fields up to *H* = 10 kOe. Some degree of irreversibility of ZFC and FCW curves was also noticed for $H \leq 10$ kOe, though the nature of the curves change with increasing magnetic field. For instance, *M/H* shows a downturn near 15 K for all the measured conditions for *H* = 100 Oe and 5 kOe, while it shows an upturn for *H* = 10 kOe.  The ZFC-FCW bifurcation starts just below the first transition for *H* = 100 Oe, whereas it sets in at the second transition for *H* = 5 and 10 kOe. These results reveal distinct effects of applied fields on these phases. However for *H* = 30 kOe, *M/H* behavior is totally different in the sense that it appears to be ferromagnetic-like and there is no difference in ZFC, FCC and FCW protocols in the entire T range of measurement.

### III B. Electrical resistivity

The temperature dependence of $\rho$ of Er$_5$Si$_3$ was measured in the absence of a magnetic field during cooling cycle and subsequent warming cycle and in the presence of magnetic fields (20 kOe and 50 kOe) during warming cycle in the *T* range 1.8-300 K. Figure 2(a) shows $\rho(T)$ (normalized to the value at 300 K) measurements; the plot is restricted to the range 1.8-60 K only, as there is no interesting feature in the measured data in the temperature range higher than 60 K. By decreasing temperature from 300 K, $\rho(T)$ is found to decrease almost linearly indicating the dominant contribution from electron-phonon scattering. At around 35 K, a sudden drop of $\rho(T)$ is observed which can be attributed to the onset of long range magnetic ordering. On further lowering the temperature, we observe an abrupt rise of $\rho(T)$ near 15 K, which is followed by a sluggish *T*-dependence. The upturn in $\rho$ below 15 K could be due to increased scattering possibly attributable to magnetic Brillouin-zone formation when the propagation vector changes. Similar behavior of $\rho(T)$ near the magnetic transition has also been reported for Dy$_7$Rh$_3$ [11] and Tb$_7$Rh$_3$ [12]. However, the transition in the present case is apparently from a gapless magnetically ordered state to a gapped ordered state, whereas in the latter case it is a transition from the paramagnetic state to a gapped AFM ordered state. With the application of magnetic field, this upturn of $\rho(T)$ vanishes (Fig. 2(a)) as though the magnetic super-zone restructured to a high-field gapless zone in the Fermi surface. The point of emphasis is that the magnetic transition around 15 K is associated with thermal hysteresis, see inset of Fig.2(a), as in the case of *M/H* data.

### III C: Heat capacity

Figure 2(b) shows the results of *C(T)* measurements in zero and in 20 kOe applied field. The zero-field data was collected for the cooling cycle as well as for the warming cycle. Distinct $\lambda$-like anomaly in *C(T)* near 35 K clearly establishes the onset of long range magnetic ordering. In addition, a very weak anomaly is noticed near 15 K, which is visible in the expanded plot shown in the inset of figure 2(b). There is a clear evidence for hysteresis in the data at this transition as in the case of $\rho(T)$. However, this transition does not result in a sharp feature at this transition in the *C(T)* data, possibly because the PPMS employing relaxation method often suppresses features due to first order transitions [13].

All these results conclusively establish that the second transition near 15 K is first-order-like, however broadened by disorder, thereby explaining the magnetic phase coexistence observed in Ref. 8.

### III D: Isothermal magnetization

In order to understand the magnetic-field-induced effects better, we have measured isothermal *M(H)* behavior at selected temperatures (see figure 3). We first discuss the *M(H)* behavior for *T* = 1.8 K. *M* increases essentially linearly with *H* initially which is not inconsistent with antiferromagnetism for the low-field values and shows a rapid increase near 12 kOe as though there is a metamagnetic transition. After this transition is complete, a weak *H*-dependence still persists. [The extrapolated saturation value turns out to be 7.5$\mu_B$/Er which is less than that expected for trivalent Er possibly because the saturation might take place at much higher magnetic fields]. While reversing the magnetic-field, a large hysteresis is observed and the virgin AFM state is not totally recoverable, unlike the magnetization behavior of the isostructural $Tb_5Si_3$ [14]. The virgin curve lies outside the envelope curve (see inset, figure 3) – a characteristic of first-order transitions. Therefore, after travelling through the transition, the high-field state partially gets arrested [10]. With increasing temperature, we find that the critical field inducing the transition in the forward cycle decreases marginally, whereas, in the reverse cycle, it increases more dramatically. Such a behavior of critical fields is a signature of kinetic arrest resulting in phase-coexistence in the intermediate-field range as well as a decrease in the area of the hysteresis loop area [10]. We note that the *M(H)* curves are completely reversible and non-hysteretic above the second transition temperature (see *M(H)* at *T* = 21 and 30 K), although field-induced spin-reorientation effects are still present in this *T* range. These results establish the distinct nature the magnetic phases below and above 15 K and also bring out signatures magnetic phase-coexistence phenomenon following broadened (field-induced) first-order transition. At temperatures higher than 35 K, *M(H)* curves show typical paramagnetic behavior.

### III E: Magnetoresistance

We now discuss the magnetoresistance anomalies brought out by the above-mentioned magnetic-field-induced transition. The results of $\rho$ measurements as a function of *H* at several temperatures are presented in Fig.4. A very complex hysteresis loop in the MR behavior is observed at 1.8 K. In the virgin curve, MR is nearly zero up to the critical field ($H_C \sim$ 12 kOe), at which it shows a sudden upturn in the field range 12 – 18 kOe followed by a decrease at higher fields. This is interesting, as usually negative MR is expected at the metamagnetic transitions. Such a positive MR has been reported by us for another compound in the same family, $Tb_5Si_3$ [14]. The critical field is the same as that obtained from *M(H)* data. We believe that such an upturn arises from inverse metamagnetism [15] in which magnetic fluctuations are induced. The drop beyond 18 kOe could be due to the suppression of these fluctuations by further application of magnetic fields with a resultant ferromagnetic alignment, though exact nature of the magnetic structure at high fields requires careful neutron diffraction studies. We also note that a large negative MR (35 %) is observed for fields higher than 30 kOe. Incidentally, positive magnetoresistance anomalies at the spin-flip transition known for $ErGa_2$, $Tb_3Co$, and $NdCu_2$ was attributed to the coexistence of different magnetic structures with enhanced scattering from the magnetic boundaries [16] and we may not rule out this possibility for the present case. In the reverse cycle (in the positive quadrant), when the field is reduced towards zero, MR behavior exhibits negligible variation till about 15 kOe with a value close to that of the high field state; subsequently there is a gradual increase with a peak close to zero, as though the critical field in the reverse direction is reduced – an observation that is consistent with *M(H)* data (see figure 3). It should be noted that the value of $\rho$ attained after returning the field to zero is intermediate between the virgin state value and the high-field value endorsing our interpretation of *M(H)* above in terms of the formation of a mixed-phase. To complete the hysteresis loop, we have also measured MR in the negative cycle (that is, second quadrant). While increasing the field in the negative cycle, after an initial dip (which could arise from a small fraction of the high-field phase in the mixed-phase), we observe a peak near 15 kOe followed by a further decrease up to *H* = 30 kOe, mimicking the behavior of virgin phase. The peak value in this quadrant is nearly half of that in the virgin curve and this is expected when there is an admixture from another phase. Comparing with the MR behavior observed in $Nd_7Rh_3$ [17], in which the transformation of the high field state gets completely arrested on returning the field to zero, remaining insensitive to further field cycling, the high field state of $Er_5Si_3$ is thus only partially arrested. Except for the virgin curve, MR behavior is found to be symmetrical for both positive and negative field cycles. A somewhat similar hysteresis loop is observed at 5 K, with some differences. That is, in the reverse cycle (in the positive quadrant), the zero-field MR is not intermediate between that of high-field phase and virgin phase, but "intermediate" between the peak value and the virgin phase value. This could imply that the phase getting partially arrested is not the very high-field phase, but an intermediate-field phase. Another difference with respect to the behavior at 1.8 K is that the magnitude at the peak in the negative quadrant is the same as that in the virgin curve. This implies complete restoration of virgin phase after returning the field to zero. The hysteresis loop area and the peak value of MR decrease with increasing $T$. Interestingly, for the data for 10K, MR value is dramatically enhanced at the peak in the reverse cycle, comparable to that in the forward cycle, as though there is a tendency to recover virgin state behavior. At higher temperatures, say at 20 K, as seen in the *M(H)* behavior, MR behavior is found to be non-hysteretic and reversible. In the paramagnetic region (e.g., at 40 and 70 K), as expected, MR is found to be negative and the magnitude decreases with increasing temperature.

The above results reveal that, at the critical field, MR is dramatically enhanced and the hysteresis and the absolute values on returning the field to zero, say at 1.8, 5 and 10 K, reveal interesting magnetic phase-co-existence phenomenon in the sense that the electrical resistivity gets accordingly modified with respect to that of the virgin phase. The phase co-existence of a high-field high-resistive and the virgin low-resistive phases observed at some temperatures (for instance, 5 K) is not very common [18]. In that sense, this compound presents an interesting situation. For comparison, the readers may see the well-known MR hysteresis loop behavior for doped $CeFe_2$ [10], $Gd_5Ge_4$ [19], $Nd_7Rh_3$ [17] and manganites [20].

### IV. Conclusions

We establish the presence of two antiferromagnetic transitions for $Er_5Si_3$, one at 35 K and the other near 15 K. We have shown that the 15K-transition arises from the disorder-broadened first-order phase transition, which is responsible for the magnetic phase-coexistence in the range 10-15 K reported earlier in Ref.8. There is a magnetic-field-induced transition in the magnetically ordered state; however, below 15 K, there is a hysteresis, both in magnetization as well as in magnetoresistance data, which becomes more prominent with decreasing temperature and an open hysteresis is distinctly observed at very low temperatures. The results reveal magnetic phase co-existence phenomenon in the *H-T* plane as well, in particular involving higher resistive high-field phase, and complex MR loops are observed. We believe that such results, as observed for $Tb_5Si_3$-related systems as well [18] point to the need to explore unusual magnetic phase coexistence phenomena. We hope that the present work will motivate further experimental and theoretical studies to understand the field-induced effects in spiral spin systems.

# References

*E-mail address: niharika@iitbbs.ac.in

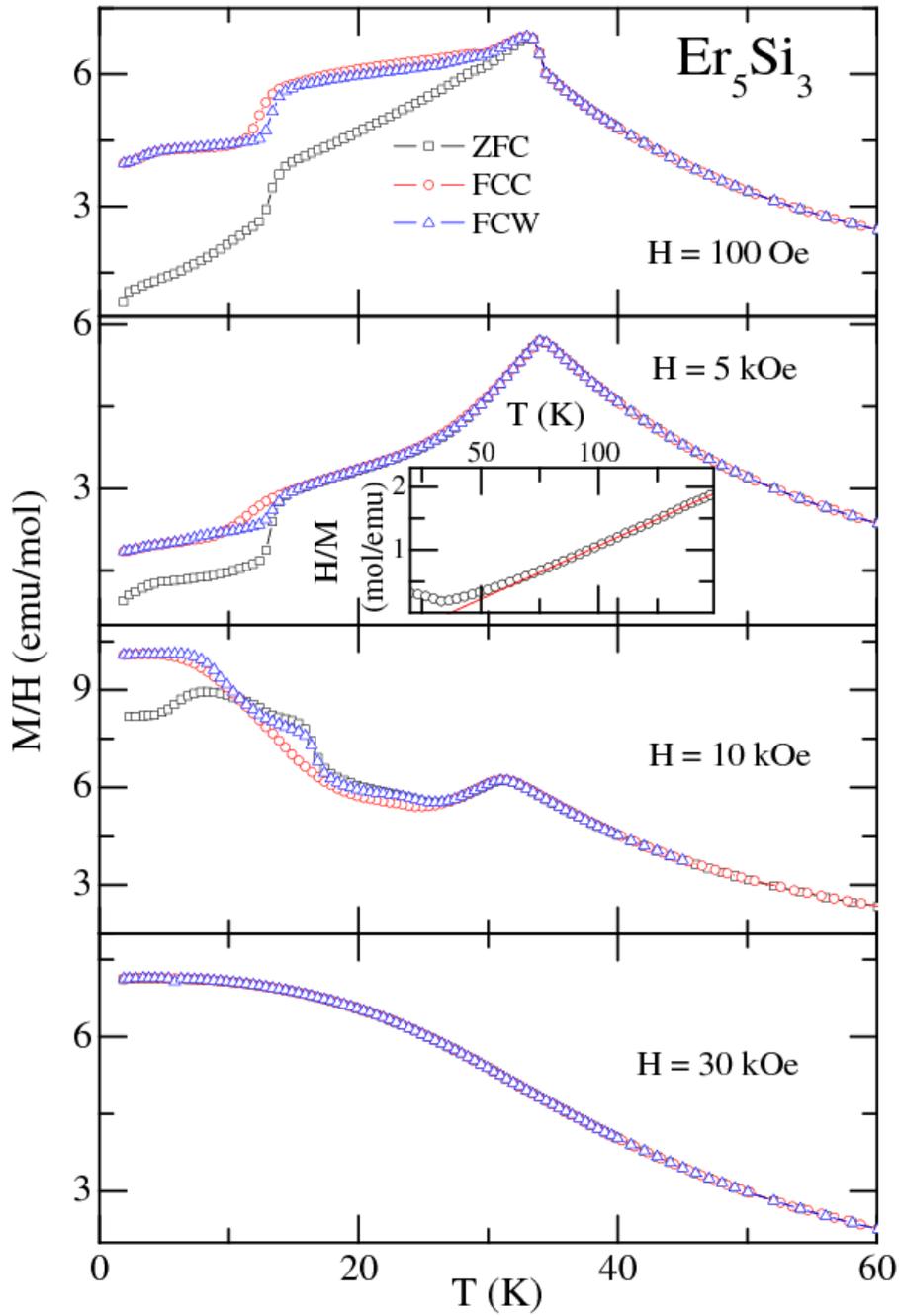

Figure 1: (color online) Dc magnetization divided by magnetic field as a function of temperature measured in presence of 100 Oe, 5 kOe, 10 kOe and 30 kOe for $Er_5Si_3$ in ZFC, FCC and FCW conditions of the specimen. Insets shows inverse susceptibility as a function of temperature and the continuous line is drawn to show Curie-Weiss region.

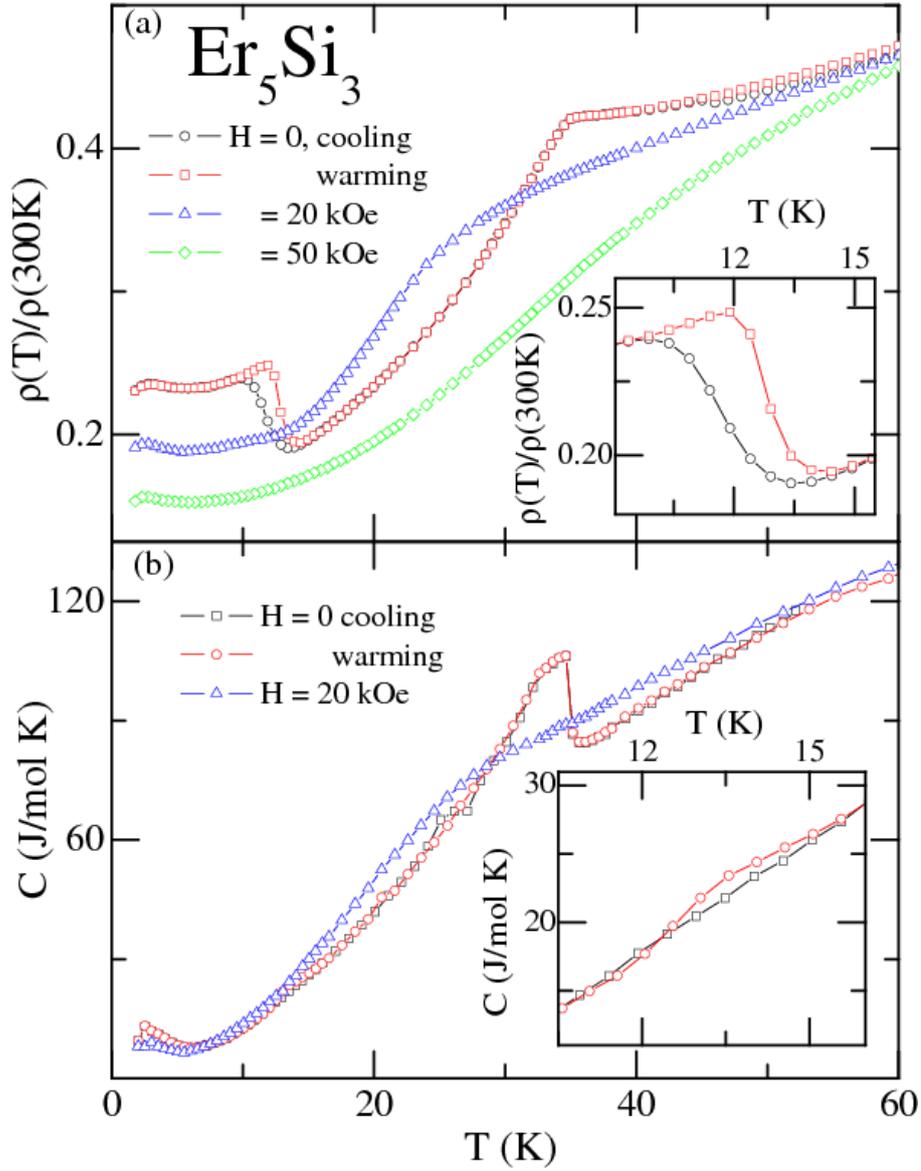

Figure 2: (color online) (a) Normalized electrical resistivity (ρ) and (b) heat capacity (C) as a function of temperature for $Er_5Si_3$ in the absence and presence of magnetic field. The zero field data was measured while cooling and warming the specimen. The thermal hysteretic behavior is highlighted in the insets. The lines through the points serve as guides to the eyes.

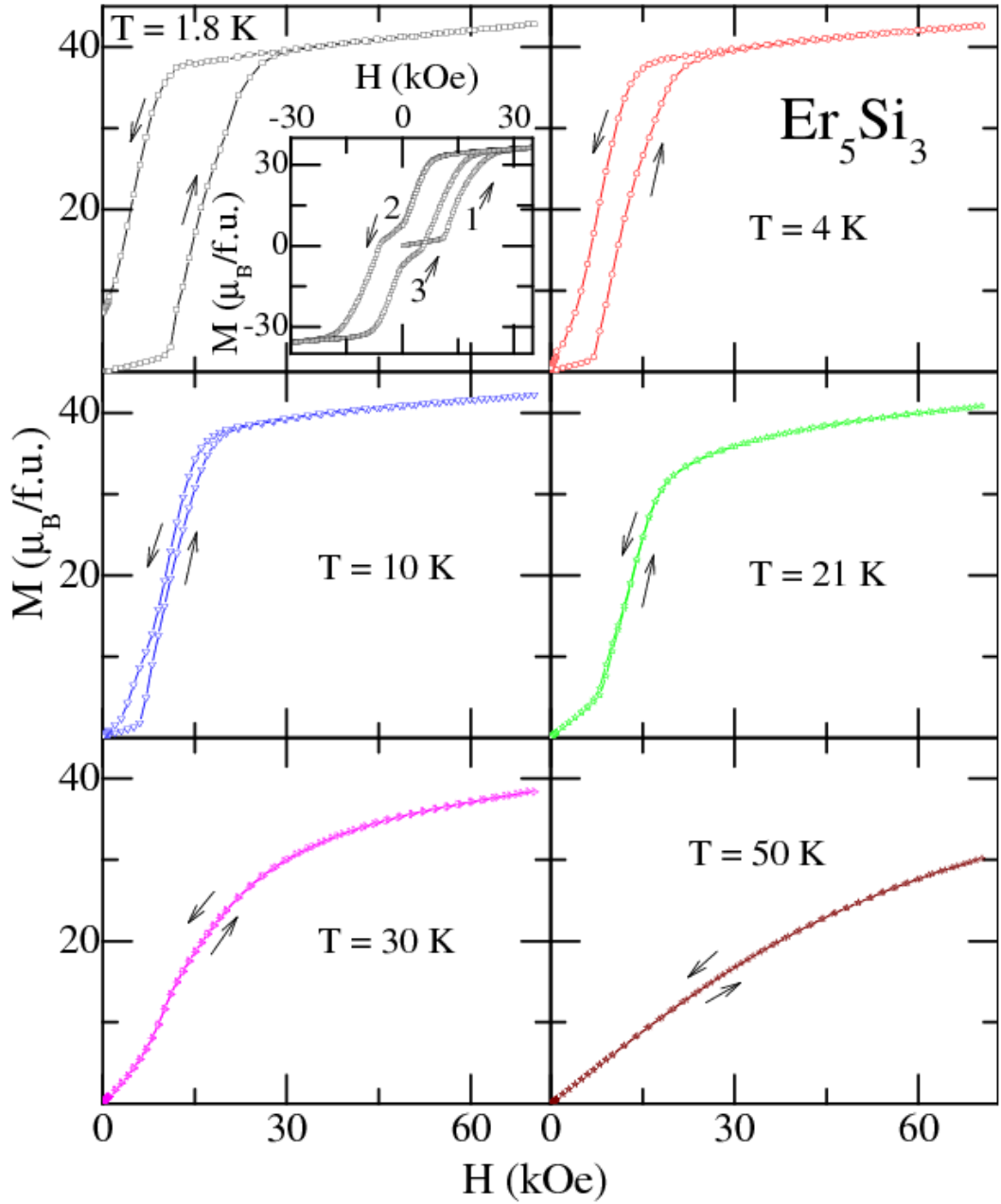

Figure 3: (color online) Isothermal dc magnetization behavior for the ZFC condition of $Er_5Si_3$ at selected temperatures measured on increasing and decreasing magnetic fields. The inset shows the hysteresis loop at 1.8 K. The lines through the points serve as guides to the eyes.

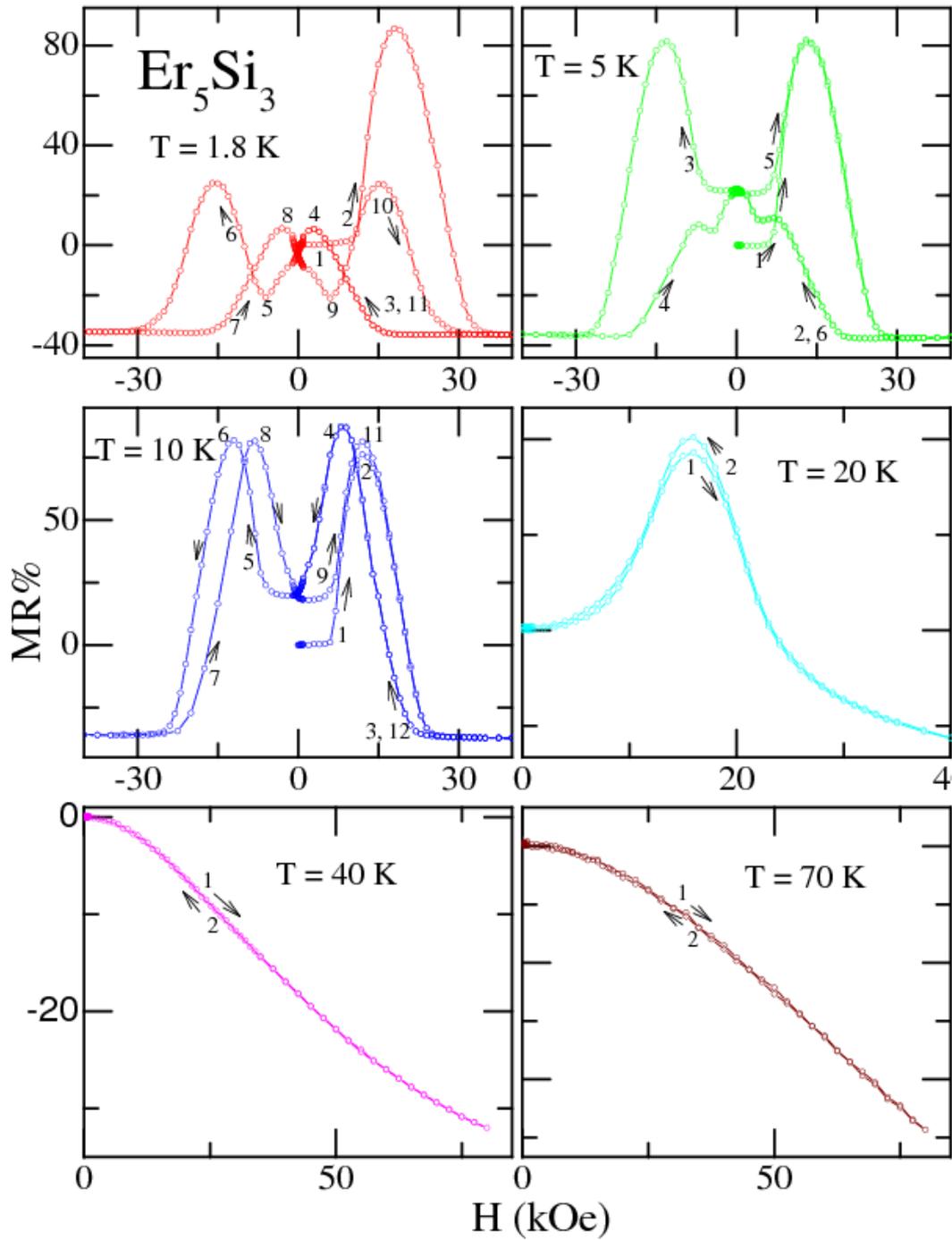

Figure 4: (color online) Magnetoresistance as a function of magnetic field for $Er_5Si_3$. Both positive and negative quadrants are shown for the temperatures below 15 K while it is restricted to only positive quadrant for T > 15 K. The numerical and arrows placed in the curves serve as guides to the eyes. The lines through the points serve as guides to the eyes.

11